\documentclass[11pt,twoside]{article}

\usepackage{asp2006}
\usepackage{epsf}
\usepackage{psfig}
\usepackage{lscape}
\usepackage{graphicx}

\begin{document}
\title{Kelvin-Helmholtz modes revealed by the transversal structure of the jet in 0836+710.}   %%% Fill in title
\author{M. Perucho and A. P. Lobanov}   %%% Fill in author names
\affil{Max-Planck-Institut f\"ur Radioastronomie, Auf dem H\"ugel, 69, 53121, Bonn, Germany.}    %%% Fill in author affiliations

\begin{abstract} %%% Abstract to run on from here.
Studying transversal structure in extragalactic jets is crucial
for understanding their physics. The Japanese led space
VLBI project VSOP has offered arguably the best opportunity for such
studies, by reaching baseline lengths of up to 36,000\,km and
resolving structures down to an angular size of 0.3 mas at 5 GHz.
VSOP observations of the jet in 0836+710 at 1.6 and 5 GHz have
enabled tracing the radial structure of the flow on scales from 2
mas to 200 mas and determining the wavelengths of individual
oscillatory modes responsible for the formation of the structure
observed. We conclude that these modes are produced by
Kelvin-Helmholtz instability in a sheared relativistic flow. Our
results point towards the stratification of the jet and the growth
of different modes at different jet radii. We also discuss the
implications of the driving frequency on the physics of the active
nucleus of the quasar.
\end{abstract}

\section{Introduction}
Resolving the transversal (or radial) structure in extragalactic
jets is a crucial step in our understanding of the physics of
these objects. On parsec scales, this has become feasible only
recently, using space VLBI\footnote{Very Long Baseline
Interferometry} observations with the VSOP\footnote{VLBI Space
Observatory Programme} \citep{lz01}. These observations revealed
the presence of a double helical structure inside the jet of
3C~273, which can be attributed to a combination of the helical
and elliptic modes of Kelvin-Helmholtz (KH) instability. Numerical
simulations further support this interpretation \citep{pe06}. We
report here on further progress of this investigation. We use VSOP
observations of the radio jet in the quasar S5\,0836+710 at 1.6
and 5\,GHz to estimate basic physical properties of the
relativistic flow, focusing specifically on the radial profiles of
its velocity and density.

The luminous quasar S5\,0836+710 at a redshift $z=2.16$ hosts a
powerful radio jet extending up to kiloparsec scales \citep{hu92}.
VLBI monitoring of the source \citep{ot98} has yielded estimates
of the bulk Lorentz factor $\gamma_\mathrm{j}=12$ and the viewing
angle $\theta_\mathrm{j}=3^\circ$ of the flow. Presence of
an instability developing in the jet is suggested by the kink
structures observed on milliarcsecond scales with ground VLBI
\citep{kr90}.

\begin{figure*}[!t]
\begin{center}
\includegraphics[height=0.40\textwidth,angle=-90,clip=true]{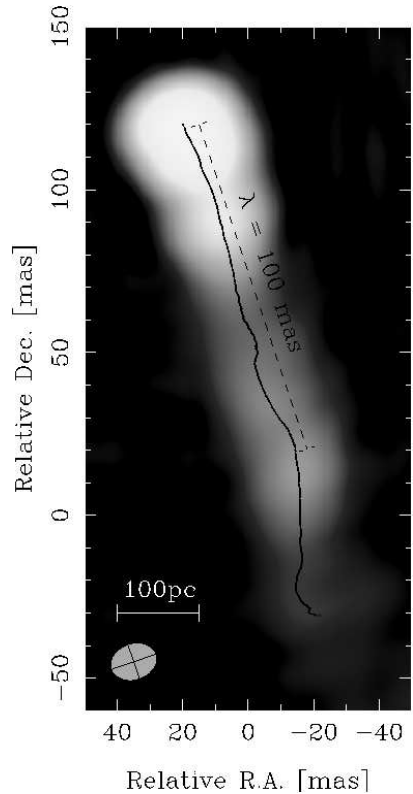}
\end{center}
\caption{1.6 GHz image of the jet in 0836+710. The ridge line of
the jet is overprinted and the 100 mas mode is identified.}
\label{fig:ridge}
\end{figure*}

In the VSOP image of 0836+710 at $5\,\rm{GHz}$, oscillations of the
jet ridge line are observed \citep{lo98}, with a wavelength of
$7.7\,\rm{mas}$. In addition to this, a $4.6\,\rm{mas}$ periodicity is
found in the the spectral index distribution. Using the vortex sheet
approximation \citep{ha00}, these structures have been
identified with the helical surface and elliptic surface modes of KH
instability, yielding estimates of basic
physical properties of the flow: the Lorentz factor
$\gamma_\mathrm{j}\sim11$, Mach number $M_\mathrm{j}\sim6$ and
jet/ambient medium density ratio
$\eta(\rho_\mathrm{j}/\rho_\mathrm{a})=0.04$. High dynamic range
VLBA\footnote{Very Long Baseline Array of National Radio Astronomy
Observatory, USA} observations of 0836+710 at $1.6\,\rm{GHz}$ show the
presence of an oscillation at a wavelength as long as
$\sim100\,\rm{mas}$ \citep{lo06}. Fig. \ref{fig:ridge} shows a
$1.6\,\rm{GHz}$ image of the jet with the ridge line of the emission
overprinted. The image also shows the identification of the 100 mas
wave. This wavelength cannot be readily reconciled with the jet parameters
determined from the two shorter wavelength oscillations, indicating
that the flow may have a complex, stratified transversal structure in
which emission at lower frequencies originates from outer layers of
the flow.

There are two basic explanations for the fact that the
longest wavelength does not fit within the scenario described above:
1)~the jet parameters derived may not be exact, or 2)~the
approximations and assumptions used in the KH linear theory are not
strictly valid. In this work we use the parameters derived in
\cite{ot98} from the spectral evolution in the jet, and relax the
approximation to vortex sheet contact discontinuity between the jet
and the ambient medium, used in the stability analysis in \cite{lo98},
exploring the possibility of the jet being sheared.

We show that the presence of a shear layer allows to fit all the
observed wavelengths within a single set of parameters, assuming
that they are produced due to KH instabilities growing in a
cylindrical outflow. In this picture, the longest mode corresponds
to a surface mode growing in the outer layers, whereas the shorter
wavelengths are identified with body modes developing in the inner
radii of the jet.

In Section~2, we describe the method used to solve the linear
stability problem for cylindric relativistic flows and provide the
respective solutions for the set of parameters given in
\cite{lo98}. These solutions are compared in Section~3 with the
wavelengths observed in the jet of 0836+710. Section~4 summarizes
the main results of this work and puts them in the broader context
of the physics of relativistic outflows.

\section{Linear Analysis}
\begin{figure*}[!t]
\includegraphics[width=0.45\textwidth,angle=0,clip=true]{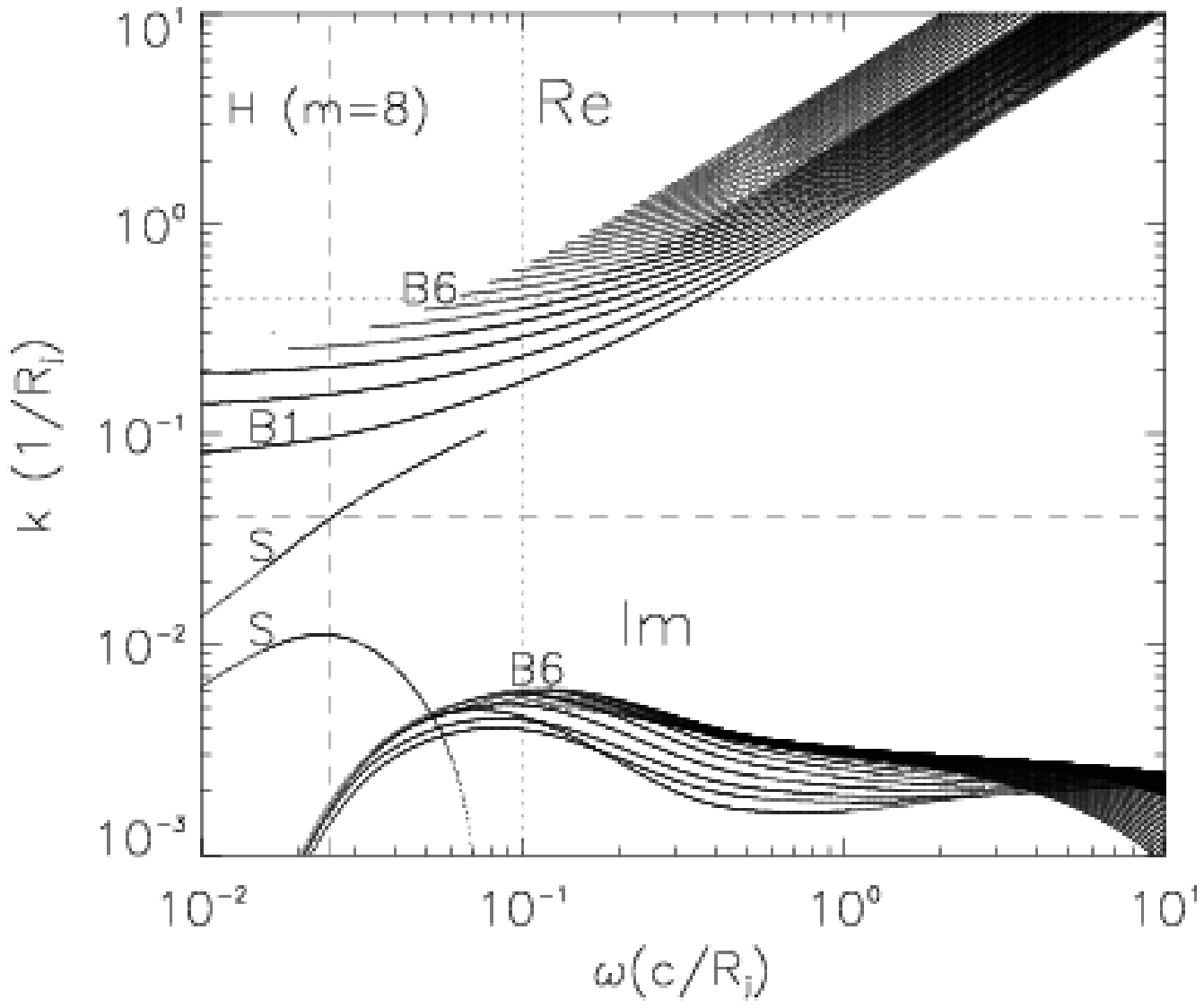}
\includegraphics[width=0.45\textwidth,angle=0,clip=true]{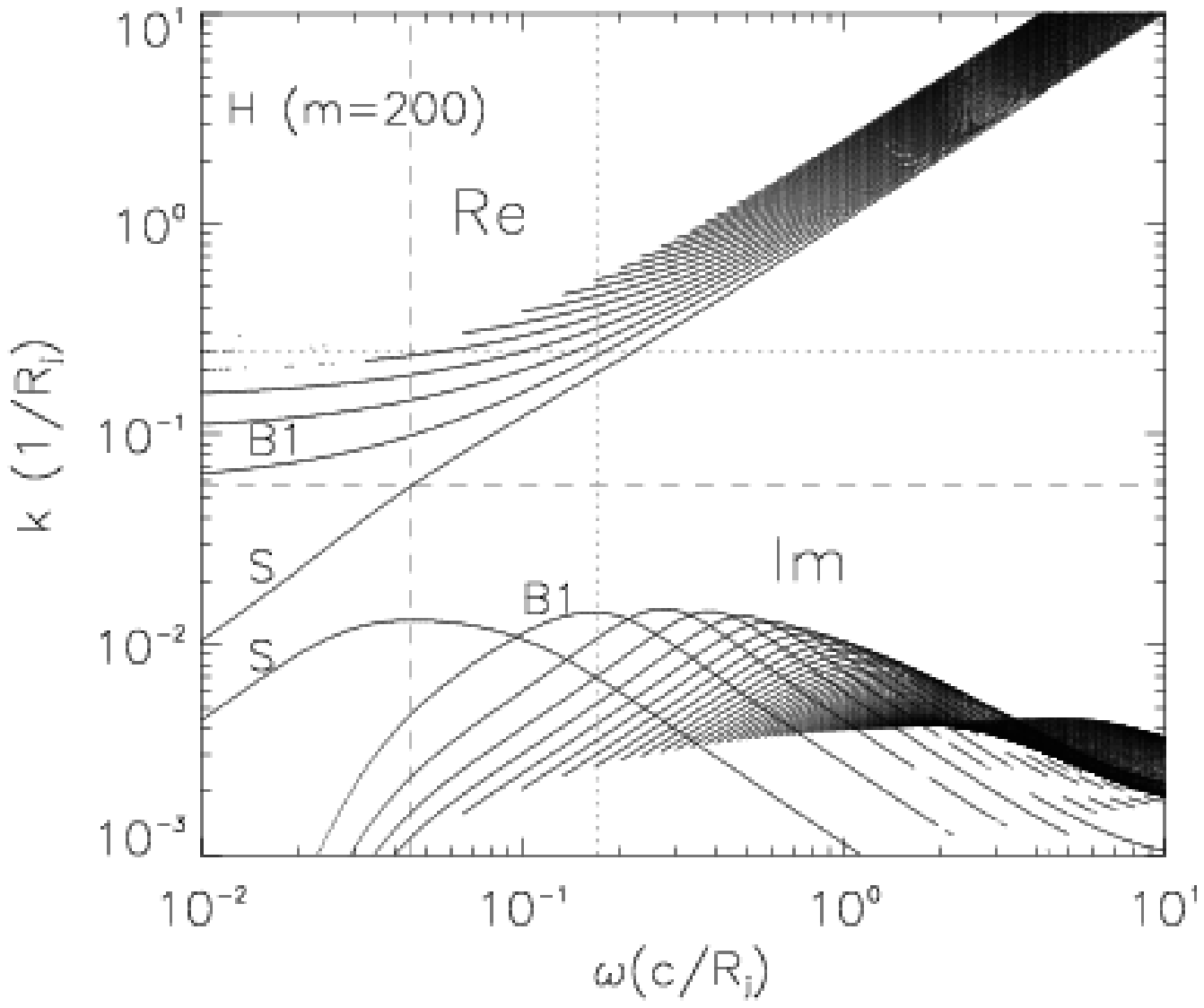}
\caption{Real (upper curves in each plot) and imaginary (lower
curves in each plot) parts of the wavenumber versus frequency for
the helical modes and the parameters given for the jet in
0836+710. The upper panel shows the solutions for $m=8$ and the
lower plot shows the solutions for $m=200$. The vertical dashed
lines indicate the minimum growth lengths of the surface modes,
whereas the vertical dotted lines indicate the minimum growth
length of the first body modes for $m=200$ and the minimum growth
length of all body modes for $m=8$. The horizontal lines indicate
the wave number at which the minimum growth length occurs. $S$
stands for the surface mode, $B1$ for the first body mode, and
$B5$-$B6$ indicate the body mode that gives the minimum growth
lengths for $m=8$.} \label{fig:sols}
\end{figure*}

We describe the flow in cylindrical coordinates ($r$, $z$,
$\phi$), with $r$ and $z$ defining the radial and axial
directions, respectively. We consider a sheared transition in the
axial velocity, $v_z$ and the rest mass density, $\rho$, between
the jet and the ambient medium of the following form (see
\citealt*{pe07} for a detailed description of the linear analysis)
$a(r) = a_\infty + (a_0-a_\infty)/\cosh (r^m)$, where $a(x)$ is
the profiled quantity ($v_z$ or $\rho$) and $a_0$, $a_\infty$ are
its values at the jet symmetry plane (at $r=0$) and at $r
\rightarrow \infty$, respectively. The integer $m$ controls the
steepness of the shear layer.  In the limit $m \rightarrow
\infty$, the configuration turns into the vortex-sheet case, as
described by \cite{ha00}. The jet and the ambient medium are in
pressure equilibrium. An adiabatic perturbation of the form
$\propto g(r)\exp [i(k_z z - \omega t)]$ is introduced in the
equations of the flow\footnote{We assume that the magnetic field
is not dynamically important.}, where $\omega$ and $k_{z}$ are the
frequency and longitudinal wavenumber of the perturbation,
respectively, and the function $g(r)$ defines the radial structure
of the perturbation. The units used are the jet radius, $R_j$, and
the speed of light, $c$. After some algebra, we obtain the
following second order differential equation for the pressure
perturbation (equation 13 in \citeauthor{Bi84} \citeyear{Bi84}):
\begin{eqnarray}
  P_1^{\prime\prime} + \left(\frac{2\gamma_0^2v_{0z}^\prime (k_z - \omega
  v_{0z})}{\omega -v_{0z}k_z} - \frac{\rho_{e,0}^\prime}{\rho_{e,0} +
  P_0}\right)P_1^\prime  + & & \label{radial-eq}\\
  \gamma_0^2\left(\frac{(\omega  -v_{0z}k_z)^2}{c_{s,0}^2} - (k_z - \omega
  v_{0z})^2\right)P_1 & = & 0. \nonumber
\end{eqnarray}
Equation \ref{radial-eq} is solved by applying the \emph{shooting
method} \citep{pr97}. This method requires integration of the
equation from the jet axis to a point outside the jet. We impose
boundary conditions on the amplitude of the pressure perturbation
and on its first derivative on the jet axis, depending on the
symmetry or antisymmetry of the perturbation on the jet axis.
Then, for a given frequency ($\omega$), a complex wavenumber is
given. The integration is done using a variable-step Runge-Kutta
method \citep{pr97}. When the integration reaches the given point
outside the jet, the value obtained is compared with the boundary
conditions there: The Sommerfeld condition, requiring the
solutions to approach zero at infinity, and the requirement of no
incoming waves at infinity. These conditions ensure that the
solutions can be given by Bessel functions inside the jet and
Hankel functions outside. Appropriate values of the complex
wavenumber that fit into those conditions are searched using the
M\"uller method \citep{pr97}. As stated above, the solutions are
obtained in the spatial domain, i.e., assuming $\omega$ is real
and $k$ is complex.  In this description, the inverse of the
imaginary part of the wavenumber, $1/{\cal Im}(k)$, gives the
growth length (or {\em e}-folding length), defined as the distance
at which the perturbation amplitude increases by one exponential
factor. We adopt the jet parameters derived in
\cite{ot98,lo98,lo06} and solve the pressure perturbation equation
with several different values of $m$. The resulting solutions are
plotted in Fig.~\ref{fig:sols} for the helical instability modes
and for two values of $m$: $m=8$ (left panels) and $m=200$ (right
panels). The $m=8$ case corresponds to a broad shear layer of
$\sim0.6\,R_\mathrm{j}$ in width, whereas the $m=200$ one implies
a shear layer width of $\sim0.1\,R_\mathrm{j}$, i.e., approaching
the vortex-sheet case.

The plots in Fig.~\ref{fig:sols} indicate that, in thicker shear
layers ($m=8$), the surface mode grows faster than the body modes.
The minimum growth lengths (maximum $k_{im}$) of the surface mode,
indicated by vertical dashed lines in Fig.~\ref{fig:sols}, are
realized at long wavelengths (horizontal dashed lines show the
wavenumber of the mode):
$\lambda^*_\mathrm{S,H,8}\sim160\,R_\mathrm{j}$, at
$\omega=0.025\,c/R_j$. The smallest growth lengths of all body
modes are achieved by the fifth and sixth order body modes at
$\lambda_\mathrm{B6,8}\sim14\,R_\mathrm{j}$.

For a thin shear layer ($m=200$) the minimum growth lengths of low
order helical body modes are similar to that of the helical
surface mode, with the minimum realized for the first body mode at
$\lambda_\mathrm{B1,200}\sim25\,R_\mathrm{j}$. The minimum growth
length of the surface modes shifts to higher frequencies and
shorter wavelengths
($\lambda^*_\mathrm{S,H,200}\sim100\,R_\mathrm{j}$) compared to
the respective values obtained for $m=8$.  For $m<8$, the growths
of all the modes are strongly reduced as the width of the shear
layer is increased. For $m>200$ the solutions converge to the
vortex-sheet case also at the highest frequencies/shortest
wavelengths.

\section{Results: From Theory to Observations}
The solutions of the linear stability problem obtained in the
previous section can be compared with the wavelengths observed in
the jet in 0836+710. This comparison requires an estimate of the
radius of the jet, in order to convert the frequencies and
wavenumbers of the solution into physical units.  The true jet
radius can be estimated from the VLBA radio maps presented in
\cite{lo98,lo06}, using the following relation: $R_\mathrm{j}
(\mathrm{mas}) = 0.5 (D_i^2-b^2)^{1/2}$, where $D_i$ is the
observed width of the jet and $b$ is the beam width transversally
to the direction of the jet. We measure the diameter of the jet
and the beamwidth at $1\%$ of the peak emission \citep{we92} at
the base of the jet, implicitly assuming that the outer parts of
the shear layer emit less than 1\% of the radio power at these
frequencies. This yields the jet radii of
$R_\mathrm{j,1.6}\sim17$\,mas and $R_\mathrm{j,5}\sim0.64$\,mas at
1.6\,GHz and 5\,GHz, respectively. If the jet is assumed to be
sheared, the radius we need to consider is that at 1.6\,GHz, as it
includes the outer, slower layers of the jet. Thus, taking the
radius of the jet to be $R_\mathrm{j}\sim17$\,mas, the $100$\,mas
structure turns into $\lambda\sim6\,R_j$, and the $7.7$\,mas into
$\lambda\sim0.2\,R_j$.

The intrinsic (rest frame) wavelengths derived from the solutions
to Eq.~\ref{radial-eq} shown in Fig.~\ref{fig:sols},
$\lambda_\mathrm{int}$, can then be transformed to the observer's
frame and expressed also in units of $R_\mathrm{j}$. The observed
wavelengths are obtained by adding the corrections for
relativistic motion, projection and cosmological effects through
the following relation:
\begin{equation}
\lambda_\mathrm{obs}=\frac{\lambda_\mathrm{int}\,\sin\theta_\mathrm{j}}
{(1+z)\, (1-v_\mathrm{w}/c\,\cos\theta_\mathrm{j})},
\end{equation}
where $\lambda_\mathrm{obs}$ is the observed wavelength, $z$ is
the redshift, $v_\mathrm{w}$ is the wave speed and
$\theta_\mathrm{j}$ is the jet angle to the line of sight. The
wave speed $v_\mathrm{w}=\omega/k$ is obtained from the solutions
to the linear problem (Fig.~\ref{fig:sols}).

We focus first in the helical surface mode, most likely
responsible for the longest observed wavelength of 100\,mas. The
wave speed of the helical surface mode ranges from
$v_\mathrm{w}\sim0.6\,c$ to $v_\mathrm{w}\sim0.8\,c$, for the
thicker and thinner shear-layer, respectively. We obtain, in the
case of the thick shear layer, a wavelength of $6.6\,R_j$, whereas
in the case of a thin shear layer we obtain $8.2\,R_j$.

For the case of the body modes, visible at 5\,GHz, we find a wave
speed of $v_\mathrm{w}\sim0.2\,c$ for the fifth and sixth body modes
(those with the fastest growths) in the case of a thick shear layer,
and a wave speed of $v_\mathrm{w}\sim0.7\,c$ for the first body mode
in the case of a thin shear layer. These modes would produce structure
with an observed wavelengths of $0.29\,R_j$ and $1.4\,R_j$,
respectively.

The wavelengths obtained for the thick shear layer are close to
those observed: $6.6\,R_j$ from theory compared to $6\,R_j$ from
the observations for the longest structure and $0.29\,R_j$ from
theory compared to $0.2\,R_j$ from the observations. This
agreement is also remarkable in the case of the $4.6$\,mas
structure, identified in \cite{lo06} as an elliptic mode and not
discussed here. This leads us to conclude that the jet is
transversally stratified (i.e., has a thick shear layer), although
the lack of error estimates precludes us from ruling out firmly
the scenario with a thin shear layer.

\section{Discussion}

The description developed in this work explains all three major
oscillations observed in the jet of 0836+710, if the flow speed
and density are both transversely stratified.  The longest
observed wavelength of 100\,mas represents the helical surface
mode, whereas the two shorter ones (7.7\,mas and 4.6\,mas) are
identified with the fifth/sixth order helical and elliptic body
modes, respectively. The former is observed in the images of the
jet at $1.6$\,GHz (see Fig.~\ref{fig:ridge}) and the latter are
observed at $5$\,GHz. The whole picture fits into the theory if
the radius of the jet at the lowest observed frequency is taken,
implying that the body modes have their maxima at inner regions of
the jet than the surface mode. If we take the radius of the jet at
$5$\,GHz, the short observed wavelengths may be interpreted as
surface modes and the longest mode cannot, in that case, be
explained by KH theory \citep{lo06}. A second implication of this
result is that the radiating particles emitting at $5\,\rm{GHz}$
occupy a central spine in a stratified jet, while the
longer-wavelength structure seen at 1.6\,GHz is generated in outer
layers of the flow.

One possible alternative to the stratified jet scenario is to
consider deviations of the basic jet parameters from those derived
in \cite{lo98,lo06}. To assess this alternative, we have
calculated the solutions of the stability equation for different
sets of parameters, varying the jet Lorentz factor, rest-mass
density ratio and specific internal energy. We conclude that the
jet in 0836+710 must have (on axis) a Lorentz factor close to 12
or slightly smaller, a density ratio of $10^{-2}$--$10^{-1}$ and a
sound speed $c_\mathrm{s,j}\sim0.2$ or slightly larger --- not
deviating much from the parameters derived in
\cite{ot98,lo98,lo06}, which supports the conclusions derived in
this paper.

Further work in the stability analysis would focus on solving the
differential equation for the pressure perturbation (Eq. 1) using a
relativistic equation of state that allows for inclusion of different
families of particles. This will enable testing the
solutions for a proton-electron jet and an
electron-positron jet, providing an useful insight into the
jet composition.

It has been shown by \cite{ha94} that the helical surface mode of KH
instability can be driven by an external periodic process. In this
case, coupling this process to the helical surface mode requires the
driving frequency to be lower than the resonant frequency. For
0836+710, the driving frequency of the helical surface mode at its
minimum growth length is $0.025\,c/R_\mathrm{j}$ would imply a driving
period of $T_\mathrm{dr}\sim5.6\cdot10^7\,\rm{yrs}$. This is similar
to the driving period of $\sim2\cdot10^7\,\rm{yrs}$ found for 3C\,449
\citep{ha94}. Periodic variations of the jet ejection axis can be
produced by a number of physical processes, including a binary black
hole and a misaligned torus as the most plausible \citep{app96}. However, a misaligned torus can only produced a period of
$10^3-10^5\,\rm{yrs}$, depending on the accretion disk and black-hole
properties \citep{lu05} in the case of 0836+710
(\cite{pl07}). On the other hand, if we combine the value derived for
$T_\mathrm{dr}$ with the estimate of the black hole mass in 0836+710
($2\cdot10^8-10^9\,\rm{M_\odot}$; \citeauthor{ta00} \citeyear{ta00}),
we obtain a companion mass of $10^4-10^7\,\rm{M_\odot}$, depending on
the separation between the two black holes
($10^{17}-10^{18}\,\rm{cm}$). The latter is thus a possible
explanation for the precession period obtained from our analysis.

Finally, the kiloparsec-scale structure of 0836+710 shows a large,
decollimated feature \citep{hu92}, without any emission between it
and the $1.6\,\rm{GHz}$ jet. New observations with high dynamic
range that have already been performed will give us more insight
into the properties of this jet on the kiloparsec scales and could
show whether the cause of the decollimation on the kiloparsec
scales is the disruption of the jet by the $100$\,mas structure,
identified here as a helical surface mode.

\begin{acknowledgements}
This work was supported in part by the Spanish Direcci\'on General
de Ense\~nanza Superior under grants AYA-2001-3490-C02 and
AYA2004-08067-C03-01. M.P. is supported by a postdoctoral
fellowship of the Generalitat Valenciana (\it{Beca Postdoctoral
d'Excel$\cdot$l\`encia}).
\end{acknowledgements}

\end{document}